# Room temperature spin injection into commercial VCSELs at non-resonant wavelengths

**Timur Almabetov[1], Petros Androvitsaneas[1], Zhongze Ren[1], Andrew Young[1], Ruth Oulton[1] and Edmund Harbord[1]**

[1] The Quantum Engineering Technology Labs, H. H. Wills Physics Laboratory and School of Electrical, Electronic and Mechanical Engineering, University of Bristol, UK, BS8 1FD
E-mail: timur.almabetov@bristol.ac.uk

### Abstract

Spin lasers leverage electron spin-polarisation to control photon polarisation, offering the potential for lower thresholds, rapid modulation, and all-optical data processing. We report successful spin injection into a commercial Vertical Cavity Surface Emitting Laser (VCSEL) using optical pumping at wavelengths of 794 nm and 810 nm. Maximum circular polarisation achieved was about 20% at 794 nm, whereas this falls to 5% with exciting at 810 nm. We attribute this to reduced spin injection due to the longer excitation wavelength in line with previous measurements on quantum wells under optical orientation. We extend the Spin-Flip Model (SFM) to account for realistic excitation conditions and find our theoretical analysis accurately reproduces experimental trends, surveying spin injection in commercial VCSELs.

## 1. Introduction

Spin lasers are semiconductor devices exploiting the fundamental quantum mechanical relationship between electron spin orientation and photon polarisation [1]. They are gaining attention due to several compelling advantages over traditional semiconductor lasers. Notably, spin lasers exhibit lower threshold currents [1,2], allowing operation at reduced power, essential for energy-efficient and sustainable technologies. Moreover, these devices enable significantly faster modulation speeds due to the inherent spin dynamics [3,4], making them ideal candidates for advanced optical communication systems. Additionally, spin lasers facilitate entirely optical data processing capabilities, harnessing the quantum spin states to manipulate and transmit information, thus paving the way for innovative computing and data transfer solutions [5-8].

Spin lasers typically function by leveraging spin-polarised carriers - electrons whose spins have a preferred orientation, and this spin polarisation is directly converted into photon polarisation during the emission process, making the emitted light polarisation-sensitive and controllable [9-11]. Such control can be achieved effectively in Vertical Cavity Surface Emitting Lasers (VCSELs), an established semiconductor laser technology renowned for their compact design, low power consumption, and ease of integration into optical circuits [12,13]. Specifically, VCSEL-based spin lasers utilize circularly polarised optical pumping to inject spin-polarised carriers efficiently [14-16].

In this paper, we demonstrate effective spin injection into commercially available VCSELs using optical pumping at two distinct wavelengths: 794 nm and 810 nm. Remarkably, successful spin injection was achieved despite the presence of polarisation-split cavity modes inherent to the device structure. Our experiments resulted in a maximum circular polarisation degree of 20% at 794 nm, significantly higher than the 5% observed at 810 nm. The enhanced spin polarisation at shorter wavelengths can be attributed to the distinct selection rules governing electronic transitions in bulk

GaAs. To validate and interpret these experimental findings, we extended the established Spin-Flip Model (SFM) [17], successfully reproducing the observed polarisation trends and elucidating underlying physical mechanisms.

## 2. Method and Results

The device used in this paper was a standard datacom VCSEL [18] in a TO-46 package with a plastic lens. To facilitate spin pumping, the plastic lens and case were physically removed, giving optical access directly to the episurface of the VCSEL. The VCSEL was mounted on a homebuilt temperature-controlled mount integrating a Peltier element and thermistor driven by a thermocontroller TCC001 from Thorlabs, maintaining the temperature with a precision <0.1 C. A Thorlabs KLD101 current source was used for electrical pumping. For optical pumping measurements, short (2.4ps) pulses were generated by a tunable diode pumped Tsunami Ti:Sapphire laser was incident on the VCSEL, focused onto the output coupler of the laser with an achromatic objective lens, focal length 7.5mm. The VCSEL emission was collected by the same lens. The pulse polarisation was set with appropriate half- and quarter-waveplates, and the emission was analysed with a separate set of waveplates, before being dispersed by Princeton Instruments TriVista spectrometer and measured on a liquid-nitrogen cooled CCD camera.

A white light source was used to measure the reflectivity of the VCSEL, normalised to a silver mirror. This normalised reflectivity is shown in figure 1(b).

The laser mode occurs at 845 nm, indicted by the narrow shallow dip in the middle of the stop band. On the short wavelength side of the stop band, there are two pronounced dips suitable for optical injection into the device. We label these as A (around 793 nm) and B (around 810 nm).

The band structure of GaAs close to the Γ point is shown figuratively in Figure 1(c). The selection rules (shown in figure 1(d)), illustrate the coupling between these two valence bands and the conduction band at the Γ point. Completely circularly polarised light exciting at the band edge couples to both spin up and spin down electrons, however, the different transition matrix element leads to a preponderance of one spin over the other, determined by the helicity and the degree of circular polarisation [9,10,19].

Transition strengths between the conduction and valence bands are determined by the selection rules [19]. These are shown in figure 1(d) for $\sigma^-$ (red) and $\sigma^+$ (blue) transitions where the HH transition is three times stronger than the light hole transition. Accordingly, 100% circularly polarised light in a bulk semiconductor leads to, at most, 50% spin polarisation. In a confined structure such as a quantum well or quantum dots, the confinement lifts the degeneracy of the heavy and light holes – if only the heavy holes are excited, 100% spin polarisation becomes possible [20,21].

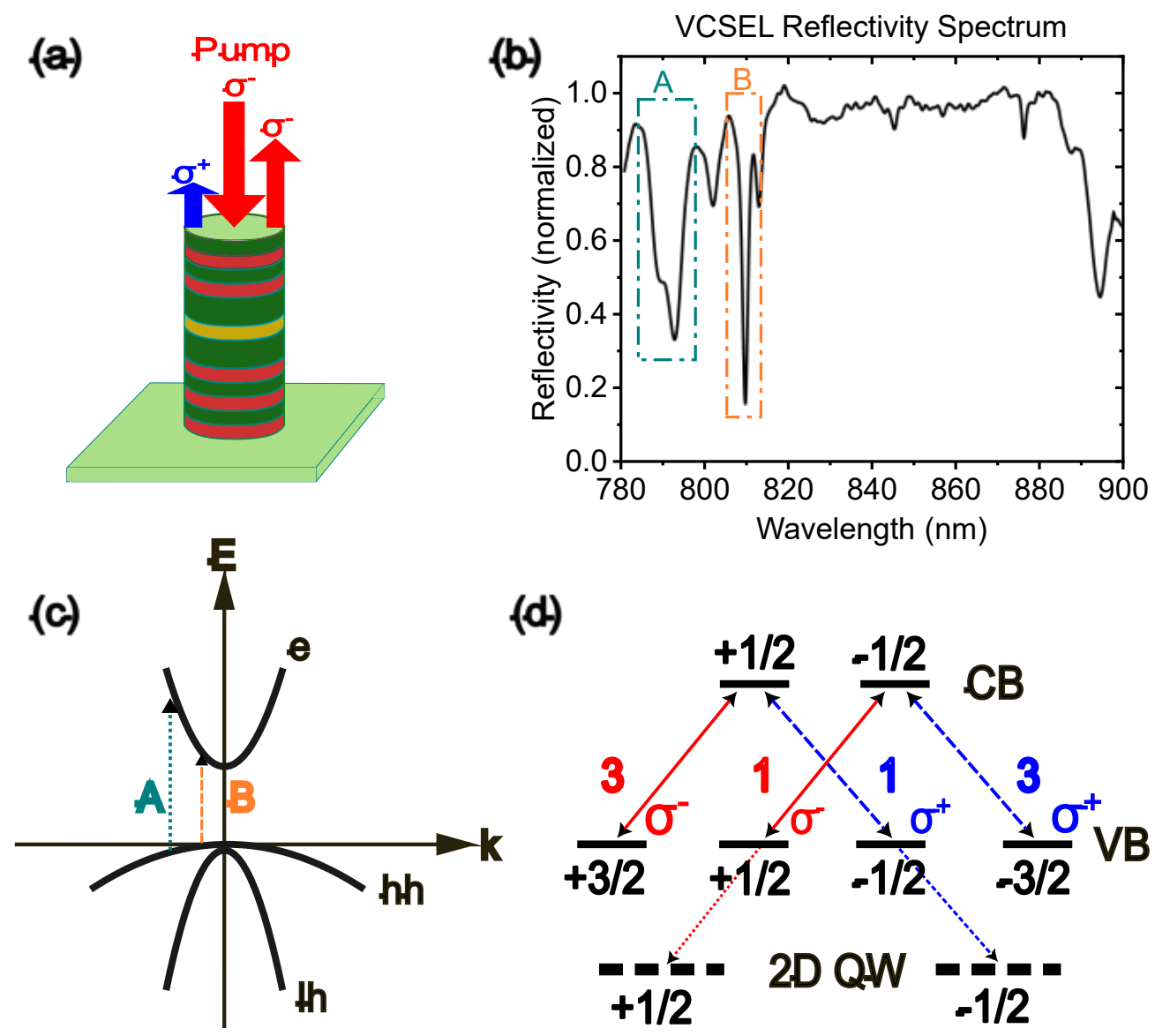

***Figure 1.*** *Illustration of optical spin-polarised pumping in a VCSEL is shown in the figure1(a). The spin-polarised optical pumping technique is concluded in $\sigma^-$ circularly polarised photons injected into the VCSEL structure, resulting in spin-polarised emission due to the connection between photon polarisation and electron spin through the optical selection rules. Figure 1(b) represents the measured reflectivity spectrum of the commercial VCSEL. On the shorter-wavelength side of the central stopband, two cavity dips suitable for optical pumping are indicated as "A" and "B". Figure 1(c) shows a GaAs band structure diagram illustrating electron transitions from heavy-hole (hh) and light-hole (lh) valence bands into the conduction band (e) under optical pumping. The different colours and transition energies denote transitions with energies corresponding to the wavelengths of A and B cavities respectively. Notably, the energy dispersion of LH band is higher at k corresponding to the transition A than to the transition B. Figure 1(d) is the detailed energy-level diagram of the selection rules in bulk and 2D quantum well (QW) GaAs [11], clarifying spin- and polarisation-dependent transitions between valence and conduction bands, highlighting allowed $\sigma^+$ (blue) and $\sigma^-$ (red) polarised optical transitions. 3:1:1:3 are relative magnitudes of the optical matrix elements.*

Under electrically pumping, the ground state of the laser is split into two linearly orthogonally polarised modes, due to the fact of VCSEL's birefringence (figure 2(a)).

Examples of optically pumped spectra are shown in figure 2(b-d), analysed in the $\sigma^+ \setminus \sigma^-$ basis. Whereas under linear polarisation there is no difference in the emission in this basis, correspond to the lack of spin injection. Under circularly polarised excitation a clear preponderance of co-polarised light can be seen, indicating successful spin lasing.

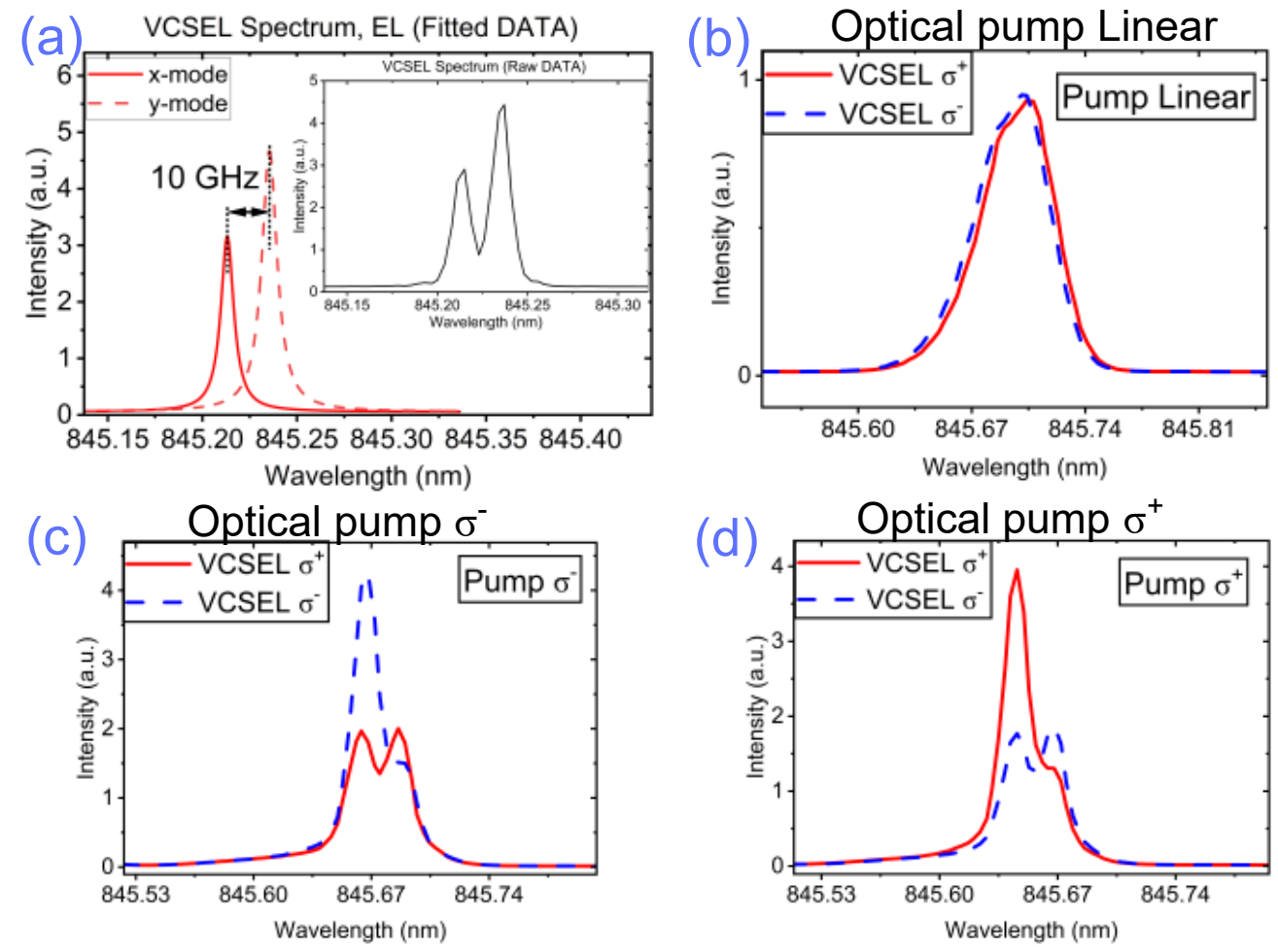

***Figure 2.*** *Figure 2(a) shows the VCSEL Spectrum (raw data in black in the inset window) under electrical pumping. The fitting of the raw data revealed the fundamental mode is split into two linearly polarised orthogonal modes (in red solid and dash lines) with frequency splitting of about 10 GHz. Figure 2(b) demonstrates two spectra measured in $\sigma^+$ (in red) and $\sigma^-$ (in blue) basis under the linearly polarised optical pumping. Figure 2(c) the signature of spin lasing is seen under $\sigma^-$ polarised pump, as the $\sigma^-$ component (in blue) dominates the $\sigma^+$ component (in red) of the VCSEL emission. Figure 2(d) shows spin-lasing under $\sigma^+$ polarised pump regime - the $\sigma^+$ component (in red) dominates the $\sigma^-$ component (in blue) of the VCSEL emission.*

This can be quantified by calculating the degree of circular polarisation, $S_3$, from:

$$S_3 = \frac{(I_{\{\sigma^+\}} - I_{\{\sigma^-\}})}{(I_{\{\sigma^+\}} + I_{\{\sigma^-\}})} \qquad (1)$$

Where $I_{\{\sigma^+\}}$ ($I_{\{\sigma^-\}}$) is the integral of the spectrum measured in $\sigma^+$ ($\sigma^-$) basis. These spin lasing curves are shown in figure 3 for injection at the two wavelengths, and the degree of polarisation resulting in lower part of the figure. As expected [2], little or no spin injection is observed before the laser threshold, whereas there is an increase in the degree of polarisation with increasing power above the threshold. In both cases, this saturates, at a value of about 20% (for excitation **A**) and 5% (for excitation **B**).

Given the circularly polarised excitation is into the GaAs, this may appear to be counter-intuitive – the closer to the bandedge, the less relaxation and according less opportunity for spin relaxation might be supposed to lead to an increase of spin-polarisation and accordingly circular polarisation. However, as excitation moves away from the Γ point (larger k), HH-LH splitting increases due to band dispersion. The light-hole band shifts upward in energy, reducing its coupling to circularly polarised excitation, effectively enhancing the heavy-hole contribution.

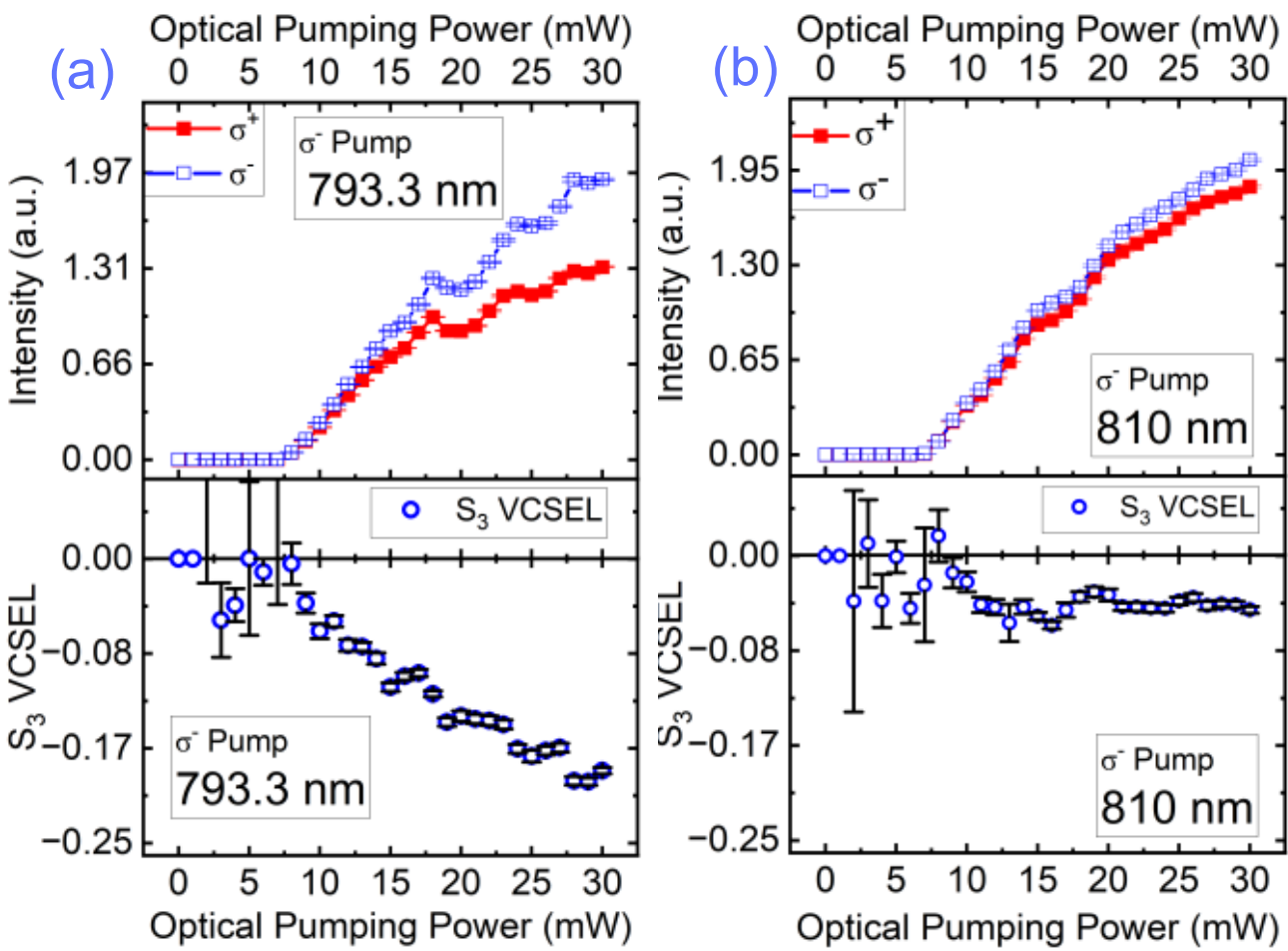

***Figure 3.*** *The figure 3(a), upper section, demonstrates L-L curves measured in $\sigma^+$ (in red) and $\sigma^-$ (in blue) basis under $\sigma^-$ polarised optical pumping at 793.3 nm, exploiting the **A** cavity dip. The spin-lasing signature becomes visible behind the threshold optical pump power, growing up to 20% of spin polarisation along with the pump power increasing. The lower section of the figure 3(a) reveals the actual $S_3$ parameter of VCSEL's overall emission. In the figure 3(b), upper section, it is seen the insufficient spin-lasing signature under $\sigma^-$ polarised optical pumping at 810 nm (**B** cavity dip), as the L-L curve measured in $\sigma^-$ basis barely prevails the L-L measured in $\sigma^+$ basis. The bottom section of the figure 3(b) numerically provides the maximum of $S_3$ parameter (around 5%) vs the optical pump power.*

## 3. Modelling

We further clarify this by applying a simple theoretical model to reproduce these trends and support our analysis. We take the well-known SFM [17] model of spin-lasing and modify it to account for realistic injection into GaAs bulk, in line with our experimental techniques.

In Figure 4(a), the spin-state are coupled together by the spin relaxation rate $\gamma_s$, and the birefringence in the device couples together the two circular polarisations at a rate $\gamma_p$. We can also include dichroism in the material $\gamma_a$, a recombination rate $\gamma$ within the device, photon lifetime $\kappa$ within the cavity, and a linewidth enhancement factor of $\alpha$.

The model has been extensively studied using numerical methods for rate equations:

$$\frac{dR_+}{dt} = \kappa(N + m - 1)R_+ - \left(\bar{\gamma}_a \cos\phi + \bar{\gamma}_p^+ \sin\phi\right)R_- \qquad (2)$$

$$\frac{dR_-}{dt} = \kappa(N - m - 1)R_- - \left(\bar{\gamma}_a \cos\phi - \bar{\gamma}_p^- \sin\phi\right)R_+ \qquad (3)$$

$$\frac{d\phi}{dt} = 2\kappa\alpha m + \left(\bar{\gamma}_p^- \frac{R_+}{R_-} - \bar{\gamma}_p^+ \frac{R_-}{R_+}\right)\cos\phi + \left(\frac{R_+}{R_-} + \frac{R_-}{R_+}\right)\bar{\gamma}_a \sin\phi \quad (4)$$

$$\frac{dN}{dt} = \gamma\left[\eta - \left(1 + {R_+}^2 + {R_-}^2\right)N - \left({R_+}^2 - {R_-}^2\right)m\right] \quad (5)$$

$$\frac{dm}{dt} = \gamma P\eta - \left[\gamma_s + \gamma\left({R_+}^2 + {R_-}^2\right)\right]m - \gamma\left({R_+}^2 - {R_-}^2\right)N \quad (6)$$

Where $\bar{\gamma}_p^+ = \gamma_p - \gamma_a \sin 2\theta$ , $\bar{\gamma}_p^- = \gamma_p + \gamma_a \sin 2\theta$ and $\bar{\gamma}_a = \gamma_a \cos 2\theta$ .

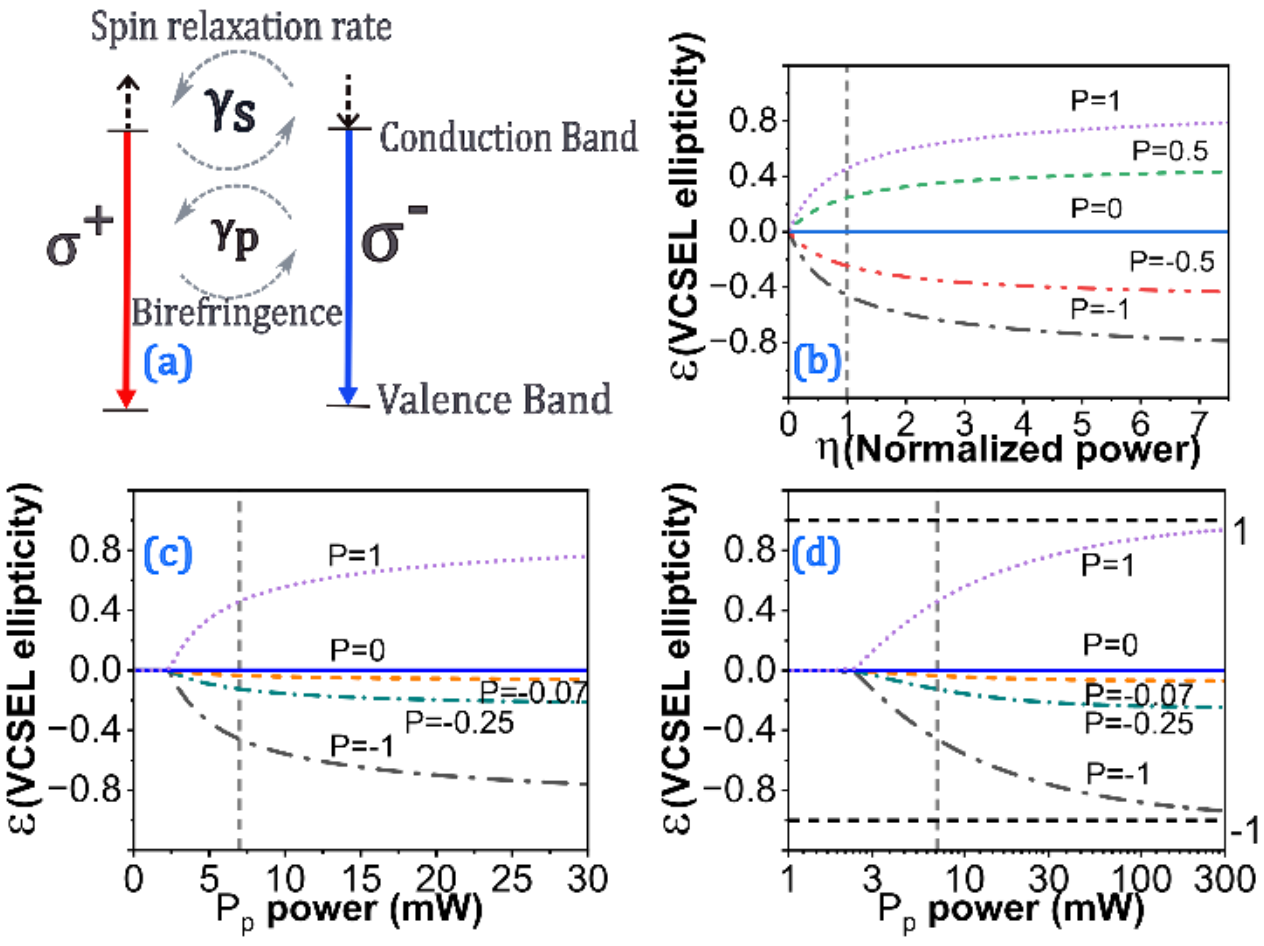

***Figure 4.*** *Figure 4(a) shows a schematic of the spin scheme in the SFM model – there are N↑ (N↓) electrons in the excited state that recombine to generate $\sigma^+$ ($\sigma^-$) light. Figure 4(b) shows the VCSEL ellipticity trend under different pump polarisation condition which is controlled by increasing normalised power $\eta$. Figure 4(c) indicates VCSEL ellipticity trend with pump power. Figure 4(d) demonstrates a comprehensive depiction of VCSEL ellipticity behaviour across high pump power regimes on a longtime scale. (The vertical grey dash line shows threshold pump power which is 7mW in these figures)*

These terms are defined in [17] as follows. The phase difference between right-circularly polarised (*RCP*, field amplitude $R_+$) and left-circularly polarised (*LCP*, field amplitude $R_-$) light is denoted by $\phi$ which can be expressed by Euler angle. The angle $\theta$ represents the misalignment between the maximum frequency axis and the maximum losses axis. Moreover, the carrier dynamics are characterised by the normalised total carrier density $N = (n^+ + n^-)$ and the difference carrier density $m = (n^+ - n^-)$ where $n^+$ and $n^-$ correspond to the spin-up and spin-down carrier densities, respectively. Meanwhile, Adams et al [17] have shown that, subject to the carrier recombination time is much greater than both the spin relaxation time and the photon lifetime, these equations could be solved algebraically for the spin laser. These assumptions are valid on our case for a VCSEL at room temperature where the spin relaxation rate ≫ 10 ns$^{-1}$.

Taking the pump polarisation [15] as $P = \frac{\eta_+ - \eta_-}{\eta_+ + \eta_-}$, and the VCSEL emission ellipticity as $\varepsilon = \frac{R_+^2 - R_-^2}{R_+^2 + R_-^2}$, we follow Adams et al [17]:

$$P = \frac{\varepsilon}{\eta}\left[\eta - 1 + \frac{\gamma_s\left(\gamma_a^2 + \gamma_p^2\right)}{\gamma\kappa\sqrt{1-\varepsilon^2}} \cdot \frac{1}{\sqrt{\left(\gamma_a + \alpha\gamma_p\right)^2 + \varepsilon^2\left(\alpha\gamma_a - \gamma_p\right)^2}}\right] \quad (7)$$

Where $\eta$ is the normalised pumping rate, which can be written in terms of the pump power [15] as:

$$\eta = 1 + \left[P_p/P_{th}\right]\left(1 + \frac{\alpha_{loss}}{g_{th}}\right) \quad (8)$$

Where $\left(1 + \frac{\alpha_{loss}}{g_{th}}\right)$ is a scaling factor, taken to be 1.5 based on physically realistic values for similar lasers [4,16,18,19].

| Parameter | Definition | Value | Reference |
|---|---|---|---|
| $\gamma$ | Carrier recombination rate | $1ns^{-1}$ | [17] |
| $\gamma_s$ | Spin relaxation rate | $50ns^{-1}$ | [4] |
| $\gamma_a$ | Dichroism | $0.2ns^{-1}$ | [17] |
| $\gamma_p$ | Birefringence | $10ns^{-1}$ | [17] |
| $\alpha$ | Linewidth enhancement factor | 2 | [22] |
| $\kappa$ | Photon field decay rate | $125ns^{-1}$ | [23] |

***Table 1.*** *lists physically realistic values for these parameters from the literature.*

We can apply this to this algebraic expression in equation (7) to adapt this model to our VCSELs. While in principle it would be possible to create a rate equation model including selection rules and spin relaxation times from an excited state, this would add further parameters to an already heavily parameterised model. Moreover, these parameters are challenging or impossible to access experimentally. Given that underlying spin relaxation and birefringence are particular to the lasing state, they will not be affected by the excitation wavelength. Accordingly, we can use P as a fitting parameter to represent the dependence on wavelength excitation. Using equation (7) we adapt to the power (normalising to 7mW as the threshold power) to observe the trend on ellipticity with power. It is clear, as shown by figure 4(d) on a logarithmic axis, that the asymptotic trend of the polarisation is to the pump polarisation P (as expected by considering equation (7) as $\eta \longrightarrow \infty$).

By discussing the selection rules under different wavelength of excitation, we can understand these results.

## 4. Discussion

The observed increase in spin polarisation at higher photon energies (793 nm versus 810 nm) appears initially counter-intuitive: after all, the higher the excitation energy the greater the energy relaxation required, which might be expected to increase the degree of spin relaxation. However, this neglects the complicated selections rules that occur at $k \neq 0$ . In bulk GaAs, the degree of optical polarisation at the band edge is found to be constant up to 330 meV above the band edge [24,25]. By contrast in quantum well, polarisation-resolved photoluminescence spectroscopy (PR-PLE) has elucidated a more complicated structure owing to the lifting of the degeneracy between the heavy holes and the light hole, due their different effective masses [10]. Excitation energy far from E-HH transition (100 meV) at low temperature (~5 K) leads to a substantial degree (~50%) of co-polarised emission. This decreases substantially as the excitation is tuned towards the transition, owing to the coupling into E-LH with the opposite selection rule, and the variation in the matrix elements associated with $k \neq 0$.

## 5. Conclusion

We have shown that it is possible to successfully inject spin into a commercially available VCSEL at two different pumping wavelengths (794nm, 810nm) in spite of the polarisation split modes in this device. This resulted in maximum polarisation of around 20% (794 nm) and around 5% (810 nm). Injecting at a shorter wavelength leads to a higher degree of circular polarisation than the longer polarisation, which can be explained by considering the selection rules under different wavelengths of excitation, in line with previously reported measurements of quantum wells. We adapted the well-known SFM model for spin lasers to account for these excitation conditions and in doing so reproduced the measured trends.

These results provide a practical entry point for polarisation-state control in standard, packaged VCSELs: under non-resonant pumping at room temperature, we obtain ~20% steady-state DCP, sufficient to deterministically bias the Stokes vector and trigger polarisation switching/excitability in off-the-shelf devices. This directly enables implementable building blocks for all-optical logic and neuromorphic photonics (spiking/RC networks) on a commercial VCSEL platform, without any epitaxial or cavity redesign [15,16,26].

## Acknowledgements

This work was supported by the EPSRC New Investigator Award "META TAMM" (EP/X029360/1) and the University of Bristol pump-prime grant "Manufacturing Future Lasers". ZR acknowledges the financial support from the China Scholarship Council (CSC) - University of Bristol joint PhD scholarship (No. 202408060152).